# Stable partial dislocation complexes in GaN as charge carrier lifetime modifiers for terahertz device applications by molecular dynamics and first-principle simulations[†]

Andrey Sarikov[1,2,3,*] and Ihor Kupchak[1,4]

[1] V. Lashkaryov Institute of Semiconductor Physics, National Academy of Sciences of Ukraine, 41 Nauky Avenue, 03028 Kyiv, Ukraine

[2] Educational Scientific Institute of High Technologies, Taras Shevchenko National University of Kyiv, 4-g Hlushkova Avenue, 03022 Kyiv, Ukraine

[3] National Technical University of Ukraine "Igor Sikorsky Kyiv Polytechnic Institute", 37 Beresteiskyi Avenue, 03056 Kyiv, Ukraine

[4] University of Rome "Tor Vergata", Via della Ricerca Scientifica 1, 00133 Rome, Italy

* Corresponding author. E-mail:sarikov@isp.kiev.ua

**Abstract**

Wurtzite GaN is a promising material for applications in photoconductive THz radiation sources. For this purpose, the photogenerated charge carriers lifetime of the order of tenths of picoseconds is required. A controllable lifetime reduction may be considered to achieve by creating recombination active stable dislocation complexes formed by mobile basal-plane Shockley partial dislocations (PDs). In this work, formation pathways and stability of PD complexes in basal planes of wurzite GaN are studied by molecular dynamics (MD) simulations. The simulations reveal the formation of stable complexes by attractive interaction of two 30° or two 90° PDs with opposite Burgers vectors located in consecutive (0001) planes. Ones formed, these complexes change neither their positions, not the atomic configurations during simulated anneal at 1500 K up to the times of 5 ns. The MD results are used as an input for density functional theory calculations to refine the atomic structures of the complex cores and to investigate their electronic properties. The calculated band structures of GaN with 30°-30° and 90°-90° dislocation complexes indicate localized energy levels in the band gap near the top of the valence band and the conduction band bottom. The calculations of the partial electronic states density confirm the possibility of electron-hole recombination between the states localized at the PD complex cores. These recombination characteristics are distinctly reflected in the calculated absorption spectra. We conclude that creating such PD complexes in required concentration may be a tool for tailoring the recombination properties of wurtzite GaN for THz radiation generation applications.

---

[†] Electronic supplementary information (ESI) available.

# 1. Introduction

Electromagnetic radiation in the terahertz (THz) frequency range (between about 0.1 and 10 THz) is promising for a vast number of practical applications including radio astronomy investigations, materials probing, medical imaging, inspecting art objects and security checks of packages and human bodies [1-9]. Use of THz radiation instead of X-ray scanning is often preferable in view of the absence of ionization-related materials damage and the need to use X-ray protected chambers in the former case.

Recent review articles [1, 3, 9] present multiple types of modern photonic and electronic THz radiation sources including optical down-converters, quantum cascade lasers, uni-travelling carrier photodiodes, photoconductors, resonant tunneling diodes, Gunn diodes, high electron mobility transistors and state-of-the-art CMOS based sources. Among them, photoconductive switches and mixers are distinguished by their simple configurations, good signal-to-noise ratio and bandwidth, and possibility of power control by device geometry and applied bias. A common material for photoconductive THz generating devices is low-temperature (200 to 300°C) grown GaAs [10, 11], which exhibits the lifetime of charge carriers in the subpicosecond range due to the large amount of carrier trapping defects, as well as high electron mobility of 8500 $cm^2/V·s$ [12]. At this, however, the generation power of the photoconductive GaAs THz sources is quite low, being limited by the thermal damage threshold of about 1 $mW/\mu m^2$ [13].

The generated THz power can be increased by substituting GaAs with wide band gap semiconductors such as ZnSe, GaN, 4H- and 6H-SiC, and $\beta$-$Ga_2O_3$ [14], capable of operation at high applied biases and temperatures. Among these semiconductors, GaN has the largest carrier mobility of ~ 1000 $cm^2/V·s$. Moreover, its direct band gap of about 3.4 eV [15] as well as high breakdown voltage of $3.3·10^6$ V/cm, high saturation velocity of $2.5·10^7$ cm/s, and high thermal conductivity of 1.3 W/cm·K [12], which more than twice exceed the respective values for GaAs, make this material an attractive alternative for THz radiation sources with much higher optical input power and operation biases as compared to the GaAs based devices.

The maximum possible frequency of the radiation generated by photoconductive THz sources is defined by the trapping time of photogenerated charge carriers. Therefore, high carrier mobility in such devices must be combined with the low carrier lifetime of the order of tenths of picosecond [1, 16]. The lifetime value may be controlled by the lifetime killing extended defects abundant in the GaN layers traditionally grown on sapphire substrates. It is known [17-24] that such layers contain various extended defect types such as threading dislocations (of the edge, screw and mixed character), stacking faults, tilted grain boundaries and inverted domain boundaries. Most of these defects enhance local carrier recombination rate by introducing deep or shallow levels into the GaN band gap. Of these, threading dislocations, with the densities up to $10^8$-$10^{10}$ cm$^{-2}$ [17, 19] (special growth methods enable decreasing this density to ~ $10^6$ cm$^{-2}$ [25]), are the most important lifetime killers in the GaN material. On the other hand, however, the lifetimes in GaN-based devices seem largely independent on the density of threading dislocations presumably due to the smaller minority carrier diffusion length compared to the inter-dislocation distances [26].

Additional control of the lifetime of photogenerated charge carriers in GaN for THz generation applications may be provided by manipulating the concentrations of gliding mobile dislocations located in {0001} basal planes. Such dislocations are easily introduced by shear stress [27] and are mobile even at room temperature having the activation energy of migration of 0.54 to 0.58 eV [28]. A number of publications report even zero activation energy for such migrations [29]. The structure and electronic properties of basal plane perfect and Shockley partial dislocations (PDs) in GaN have been studied by ab initio calculations [17, 23, 24]. It has been demonstrated [23] that perfect basal plane dislocations ($\vec{b} = \frac{1}{3}\langle 11\bar{2}0 \rangle$ [19]) with N-polarity of the dislocation core introduce shallow states into the band gap, while the dislocations with Ga-polarity introduce both shallow and deep levels. Such dislocations dissociate into pairs of Shockley partials with the Burgers vectors forming the angles 30° and 90° with the dislocation lines, according to the following reaction [19]:

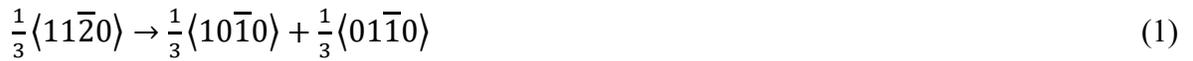
$$\tfrac{1}{3}\langle 11\bar{2}0 \rangle \rightarrow \tfrac{1}{3}\langle 10\bar{1}0 \rangle + \tfrac{1}{3}\langle 01\bar{1}0 \rangle \qquad (1)$$

Density functional theory calculations [17, 24] have demonstrated that an N-core 90° Shockley partial dislocation introduces a donor level at $E_v$ + 0.87 eV, which may be responsible for the absorption peak at 2.4 eV, while a Ga-core dislocation hosts an acceptor level at $E_v$ + 1.11 eV, which may contribute to the GaN yellow luminescence. It has been also shown [17] that Ga-core 30° Shockley partial dislocations exhibit a localized filled band ranging from $E_v$ + 0.93 eV to $E_v$ + 1.83 eV, while N-core dislocations have an empty deep band between $E_v$ + 0.66 eV at the Brillouin zone boundary to $E_v$ + 2.18 eV at the $\Gamma$-point.

Easy movement and multiplication of basal plane Shockley partial dislocations under the action of a shear stress or due to recombination of non-equilibrium charge carriers (radiation enhanced dislocation glide [30]) make them hardly suitable to control the lifetime of charge carriers in GaN-based photoconductive THz radiation sources. Our previous experience in particular with cubic Si carbide (3C-SiC) [31-33] and hexagonal Si and Ge [34] evidences that Shockley partial dislocations can form double and triple dislocation complexes upon their interaction. Once formed, these complexes change neither their positions nor atomic configurations during subsequent processing of the material or operation of the device based on it. Moreover, such complexes may introduce energy levels into the band gap thus promoting recombination of charge carriers and decreasing their lifetime [32]. Hence, manipulating Shockley partial dislocations in GaN by creating (possibly) recombination active stable dislocation complexes may be promising for control of the lifetime of charge carriers in this material for THz generation applications.

In this work, we use molecular dynamics (MD) simulations to study the interaction of basal plane 30° and 90° Shockley partial dislocations in wurtzite GaN. We demonstrate that such interaction may lead to formation of stable double dislocation complexes with zero Burgers vectors. Using ab initio calculations, we show that such complexes introduce energy levels into the material band gap thus enhancing recombination of non-equilibrium charge carriers. We conclude that creating stable dislocation complexes with zero Burgers vector may be considered for control of the lifetime of charge carriers in wurtzite GaN for THz radiation generation applications.

## 2. Methods

### 2.1. MD simulations

The formation dynamics and stability of partial dislocation complexes in wurtzite GaN were studied by molecular dynamics simulations using Large-scale Atomic/Molecular Massively Parallel Simulator (LAMMPS) [35]. The interaction forces between the atoms in the considered systems were calculated based on the Stillinger-Weber potential [36] implemented in a standard LAMMPS package. This potential was demonstrated before to provide a reliable description of the characteristics and properties of various dislocation types in wurtzite GaN [36, 37]. All the MD simulations were carried out in a Nose-Hoover barostat regime (NPT ensemble) applying periodic boundary conditions to the simulation cells in all directions. The simulation time step was set to 2 fs based on energy conservation of the cells in testing simulation runs. The simulation temperature was 1500 K, which is well below the GaN melting temperature (~2800 K [38]) to avoid pre-melting cell perturbations on the one hand and high enough to ensure reasonably fast dislocation dynamics on the other hand. The configurations of the simulation cells at different time steps were analyzed using the Open Visualization Tool (OVITO) software [39].

A ca. 334×236×10 Å$^3$ rectangular block of wurtzite GaN containing 64800 atoms and shaped by the crystallographic planes (10$\bar{1}$0), (0001) and (1$\bar{2}$10) was taken as a primary cell. Such cell dimensions were sufficiently large so that the dislocation dynamics was not affected by interactions with the image dislocations appearing due to periodic boundary conditions applied to the cells. To prepare the simulation cells, Shockley partial dislocations with the Burgers vectors at 30° or 90° with respect to the dislocation lines (hereafter called 30° and 90° Shockley partial dislocations, respectively) were inserted into the primary cell by displacing all the cell atoms by the vectors **u**($u_x$, $u_y$, $u_z$), the components of which were calculated as follows [40]:

$$u_x = \frac{b_{edge}}{2\pi}\left[\arctan\frac{y}{x} + \frac{x \cdot y}{2(1-\nu)\cdot(x^2+y^2)}\right] \quad (2a)$$

$$u_y = \frac{-b_{edge}}{2\pi}\left[\frac{1-2\nu}{4(1-\nu)}\ln(x^2+y^2) + \frac{x^2-y^2}{4(1-\nu)\cdot(x^2+y^2)}\right] \quad (2b)$$

$$u_z = \frac{b_{screw}}{2\pi}\arctan\frac{y}{x} \quad (2c)$$

Here, $b_{edge}$ and $b_{screw}$ are the edge and the screw components of the dislocation Burgers vector, respectively, $\nu = 0.183$ is the GaN Poisson ratio [41], and $x$, $y$ and $z$ are the atom coordinates, respectively. It should be noted that the expressions (2a-2c) apply to the coordinate system where the Z-axis is directed along the dislocation line and the X-axis is located in the dislocation slip plane [42].

Two types of simulation cells, namely the one with two dipoles consisting of 30° Shockley partial dislocations and the other one with two dipoles consisting of 90° partial dislocations located in neighboring basal planes were prepared. The initial dislocation configurations of both simulation cell types are shown in Figs. 1(a) and (b). Each dislocation dipole consisted of the dislocations with opposite Burgers vectors separated by stacking faults naturally formed in the basal planes as a result of inserting the dislocations. The dislocation line directions were [1$\bar{2}$10], which are typical for this type of partial dislocations in wurtzite-lattice crystals [34]. The distances between the dislocations in the dislocation dipoles were equal to the half of the cell dimension in the [10$\bar{1}$0] direction. The dipoles were inserted into the neighboring basal planes with a shift relative to each other by a half of the stacking fault length to ensure equidistant dislocations distribution. In view of the geometry of the wurtzite crystallographic structure of the GaN cells, the Burgers vectors of the leftmost and the rightmost dislocations in the dipoles were also opposite. In such configurations of the initial simulation cells, attraction of the leftmost and rightmost dislocations from the two dislocation dipoles and formation of double dislocation complexes with zero total Burgers vectors is anticipated. Our previous experience with basal plane defects in hexagonal Si and Ge core-shell nanowires [34] evidences that such complexes might be stable. Moreover, stable dislocation complexes may introduce energy levels into the material band gap [32] thus providing a tool for modifying the charge carrier lifetime.

*2.2. DFT calculations*

Density Functional Theory (DFT) calculations were used to unveil the effect of stable dislocation complexes on the GaN energy band structure. The model used periodic supercells containing pairs of dislocation complexes with the initial atomic positions provided by the molecular dynamics simulation results. A supercell was orthorhombic, obtained by cutting out the bulk region around the defects in typical armchair-zigzag approach, with the dimensions of 144.33×15.85×3.21Å$^3$ and contained 624 atoms. The studied systems contained pairs of complexes formed by 30°-30° and 90°-90° dislocation pairs, separated by double-plane stacking faults, and one pristine bulk GaN as a reference. Electronic structure calculations were done by DFT [43] in generalized-gradient approximation (GGA), as implemented in the Quantum Espresso package [44]. Perdew-Burke-Ernzerhof (PBE) pseudopotentials [45], which include 13 valence electrons for Ga and 5 valence electrons for N, were chosen. A 60 Ry cutoff for the smooth part of the wave function expansion and 500 Ry for the augmented charge density were set to ensure convergence of the results. Integration of the Brillouin zone was performed using a 1×8×2 k-points mesh centered on Γ-point, generated by the Monkhorst-Pack scheme [46], and Methfessel-Paxton smearing [47] with the parameter of 0.005 Ry. All the studied systems were relaxed over all the internal coordinates until the Hellmann-Feynman forces dropped below $10^{-4}$ a.u. After that, the electronic (band) structure was calculated. Well-converged band-structure calculation for primitive 2-atom GaN bulk showed that our approach underestimated the band gap: the value of 1.93 eV was obtained, while the experimental value is 3.4 eV [15]. Thus, a scissor operator of 1.47 eV was introduced into the band structure calculations to compensate this inconsistency. It is worth noting that in the case of 90°-90° dislocation complexes, the geometry optimization generated an atomic configuration turning the system into metallic state at some steps of the standard BFGS algorithm [48]. Therefore, we carefully repeated the calculations for both systems to ensure the finding of the ground state.

## 3. Results and discussion

### *3.1. Formation of stable 30°-30° and 90°-90° partial dislocation complexes*

The evolution of the cell with 30° Shockley partial dislocations during the MD simulated anneal is shown in Fig. 2. Fig. 1 S. I. shows the respective evolution of the simulation cell with 90° dislocations. As can be seen from these figures, as expected, attractive interaction between the dislocations with the opposite Burgers vectors, located in neighboring basal planes, leads to formation of double dislocation complexes. Ones formed, these complexes change neither atomic configurations nor positions in subsequent simulations up to the time of 5 ns at 1500 K. This allows us to conclude that the obtained complexes are stable. As can be further seen from Fig. 2, the inner 30° dislocations, which have a Ga core, were immobile in our simulations, while the evolution of the entire defect structure of the simulation cell was due to the motion of the N-core dislocations. These findings have a full agreement with the experimental data of [30], where high-resolution transmission electron microscopy demonstrated a much higher mobility of the N-core Shockley partial dislocations during radiation enhanced dislocation glide as compared to that of the Ga-core dislocations.

In the case of the 90°-90° dislocation complexes, the mobility of the Ga-core 90° dislocations (inner ones), being different from zero (see Fig. 1 S. I.), was nonetheless substantially lower as compared to that of the N-core dislocations, similar to the 30° dislocations case, again with agreement with [30]. Therefore, the formation kinetics of the 90°-90° partial dislocation complexes was also primarily defined by the motion of the N-core dislocations. It should be also noted that the mobility of the N-core 30° Shockley partial dislocations in GaN exceeded that of both the Ga- and N-core 90° dislocations in our MD simulations. This led to the much smaller simulation time needed for the formation of the stable 30°-30° complexes as compared to that for the formation of 90°-90° complexes (~300 ps versus ~4.2 ns, respectively).

As can be further seen from the insets in panels (d) of Fig. 2 and Fig. 1 S. I, both 30°-30° and 90°-90° dislocation complexes formed have two possible atomic configurations depending on the dislocation stacking order. We refer to these configurations viewed along the dislocation lines, as the flower-like (for the right 30°-30° and left 90°-90° complexes) and butterfly-like (for the left 30°-30° and right 90°-90° complexes) ones. No perfect flower-like configuration of the 90°-90° dislocation complex could be obtained in the MD simulations (see Fig. 1(d) S. I.). However, the perfect configuration of this complex is retrieved by the ab initio optimization of the simulation cell structure as discussed in the next section (see Fig. 2 S. I. (b)). The stability properties, i.e. no change of both the position and atomic configuration of the complex, equally correspond to all the four obtained complex types.

The volumetric strain as well as the elastic energy maps of the initial simulation cells and the cells after the formation of the stable dislocation complexes for the cells with 30° and 90° Shockley partial dislocations presented in Fig. 3 and Fig. 3 S. I., respectively, allow one to understand the mechanisms of the formation and stability of the dislocation complexes obtained in the MD simulations. As can be seen from these figures, both the elastic strain and the elastic energy in the lattice substantially decrease upon formation of the complexes with the zero value of the Burgers vectors. This situation is easily understood because the elastic energy introduced by a dislocation core is roughly proportional to **b**$^2$, **b** being the dislocation Burgers vector [40]. This finding also agrees with the results of our earlier publications [32, 49], where zero-Burgers vector dislocation complexes composed by three Shockley partial dislocations introduced hardly any excess elastic energy and strain into the crystal lattice of a 3C-SiC crystal. As one can further see from Fig. 3 and Fig. 3 S. I., the elastic stress and the elastic energy introduced by the left dislocation complexes in both cases have higher values as compared to those of the right dislocation complexes, for which they are almost zero. Lower introduced energy should promote easier formation of the right dislocation complexes as compared to the left ones, in real materials structures. This finding has an agreement with the experimental results obtained for hexagonal-phase Si nanowires, for which formation of numerous 30°-30° basal plane dislocation complexes with the flower-like configuration was observed, while no complexes with the butterfly-like configuration was detected [34].

*3.2. Effect of stable dislocation complexes on the band structure of wurtzite GaN*

The band structure calculated for the supercells containing 30°-30° and 90°-90° dislocation complexes is shown in the central and right panel of Fig. 4, respectively. For comparison, the band structure of the initial (defect-free) GaN calculated for the same supercell as the defective ones is also presented in Fig. 4 (left panel). As can be seen from this figure, both types of dislocation complexes induce the appearance of local levels in the band gap of GaN. The wave functions corresponding to the minimum and maximum of the defect zones (the highest occupied molecular orbital, HOMO, and the lowest unoccupied molecular orbital, LUMO levels) are shown in Fig. 5. As can be seen from this figure (see upper panel), the HOMO (blue) and LUMO (red) levels for the 30°-30° dislocation complexes are formed by the states of the atoms of the same complex with flower-like atomic configuration. The energy of the HOMO level at the point D is 1.45 eV and that of the LUMO level is 3.02 eV.

For the 90°-90° dislocation complexes, the quantity of the local levels is higher as compared to the 30°-30° dislocations case. Moreover, these levels are located closer to the edges of the valence and conduction bands. As can be also seen from Fig. 5 (lower panel), the HOMO and LUMO levels are spatially separated. In particular, the former are formed by the states of the atoms at the flower-like 90°-90° dislocation complex, while the latter correspond to the complex with butterfly-like atomic configuration. We have to mention here that in view of the necessity to preserve periodic boundary conditions of the supercells, we cannot separate the left and right dislocation complexes to study their individual effect on the energy levels of wurtzite GaN and must only consider their combined influence.

For a more detailed analysis of the local levels, we calculated the partial density of states (PDOS). Fig. 6(a) shows the PDOS for the defect-free GaN along the $X$-axis. In fact, this is the partial density of electronic states belonging to the atoms in a thin $YZ$ layer of the supercell for a given $X$ coordinate. The intensity of the black color corresponds to the density of states value in relative units. The dotted lines indicate the band gap. As can be seen from this figure, the PDOS of the defect-free GaN is homogeneous along the $X$-axis, as expected for an ideal crystal structure. The total density of states shown in the vertical inset, which reflects the PDOS integrated along the $X$-axis, is also fully consistent with the ideal crystal structure. It should be noted that the PDOS is very small near the bottom of the conduction band and practically invisible in the figure, but clearly visible in the full density of states in the vertical tab.

In contrast to the defect-free crystal, the PDOS of the supercell with 30°-30° dislocation complexes shows certain inhomogeneities, which are clearly visible near the edge of the conduction band and also deep in the valence band (see Fig. 6(b)). This is especially evident in the conduction band in the energy region of 5-6 eV and in the vicinity of $X = 40$ Å, which corresponds to the butterfly-like dislocation complex (see the atomic configuration below). Moreover, a discrete level with an energy of about 0.1 eV is also clearly visible here. The other dislocation complex having the flower-like atomic configuration and located at about $X = 105$ Å has a much more complicated local energy structure. Here, a PDOS redistribution from the valence and conduction bands to the local levels in the band gap is observed. As a result, a series of PDOS maxima appear near the edge of the conduction band, as well as in the band gap in the energy range up to 1.8 eV. These maxima are reflected in the integral DOS as well as the changes in the shape of the edges of the valence and conduction bands.

The situation is different for the supercell with the 90°-90° dislocation complexes (see Fig. 6(c)). Here, the inhomogeneities are more pronounced and the redistribution of the density of states occurs not only in the vicinity of the complexes, but in a wider region. Thus, the maximum density of states in the valence band between 0 and 60 Å "bends" deep into the band. At the same time, a separate maximum at the very edge of this band is clearly visible in the vicinity of the dislocation complex at $X = 40$ Å having the flower-like atomic configuration. The same DOS maxima are also observed near the complex at $X = 105$ Å with the butterfly-like atomic configuration, but the bending of the valence band in this case is in the opposite direction, and the series of maxima are more widespread deeper into the band gap.

Fig. 7 shows the absorption spectra calculated for the GaN supercell with the 30°-30° and 90°-90° dislocation complexes. These spectra are averaged over all the $(X, Y, Z)$ directions and smoothed by applying a Gaussian convolution with a 25 meV broadening. The dotted line corresponds to the band gap of GaN. As can be seen from this figure, the initial GaN has a rather sharp edge of fundamental absorption (black line), and there are no bands in the lower energy region. In the case of the 30°-30° dislocation complexes (red line), the fundamental absorption edge becomes flatter and shifts towards lower energies by ~ 0.25 eV. Moreover, weakly pronounced broad absorption bands appear in the energy range of 1.5-2.5 eV.

For the supercell with the 90°-90° dislocation complexes, the situation is much more complicated. Here, the fundamental absorption edge is even more shifted towards lower energies, it is rather flat, and the spectrum itself shows a large number of well-defined bands with the intensities much higher than in the case of the 30°-30° complexes. Obviously, the higher intensity of these bands is associated with a higher density of electronic states, both local defect levels in the band gap and the states at the band edges. Although electron-hole recombination occurs within the complex in the 30°-30° case, while for the 90°-90° complexes it occurs between different complexes, these localized states behave differently during recombination processes, which is evident from the calculated absorption spectra. Recombination of carriers between the states with a higher density will be faster, which leads to an increase in the absorption coefficient in the corresponding energy region.

## 4. Conclusion

In conclusion, in this work, formation pathways and stability of Shockley partial dislocation complexes in basal planes of wurzite GaN were studied by molecular dynamics simulations. MD simulations were performed using the LAMMPS software with the Stillinger-Weber potential describing the Ga and N atom interaction. The simulation results revealed that interaction of pairs of 30° or 90° PDs with opposite Burgers vectors situated in consecutive (0001) planes leads to formation of double dislocation complexes with zero total Burgers vectors. Once formed, these complexes changed neither the positions, not the atomic configurations of the cores during simulated anneal at 1500 K up to the times of 5 ns. The complex stability originates from the decrease of mechanical strain and the dislocation strain energy of wurtzite GaN as a result of complex formation.

Density functional theory calculations within the generalized gradient approximation were used to investigate the electronic structures and refine the atomic configurations of the formed PD complexes. The initial atom positions for the DFT calculations were taken from the MD simulation output. The calculated band structure of GaN containing the dislocation complexes indicates the appearance of the localized energy levels in the band gap near the valence band top and conduction band bottom. Analysis of the wave functions of these states shows their association with the atoms at both the same and different poles of the dislocation complex dipoles. The computed partial density of electronic states confirmed a possibility of electron-hole recombination between the states localized at the complex dipole poles. At this, for the 30° PD complexes, the recombination occurs within a single pole, while for the 90° PD complexes, it occurs between different poles. These recombination process characteristics are distinctly reflected in the calculated absorption spectra showing an extension in the low-energy region. It may be concluded therefore that creation of recombination-active stable double dislocation complexes in basal planes of wurtzite GaN by manipulating mobile Shockley partial dislocations may be considered as a tool for controlling the lifetime of non-equilibrium charge carriers for applications in photoconductive THz radiation generators.

**Conflicts of interest**

There are no conflicts to declare.

**Data availability statement**

The data for this article is contained within the article text as well as in the electronic supplementary information.

**Acknowledgement**

The authors acknowledge support of their research by the Alexander von Humboldt Foundation in the framework of the Digital Cooperation Fellowships for Alumni in Ukraine program.

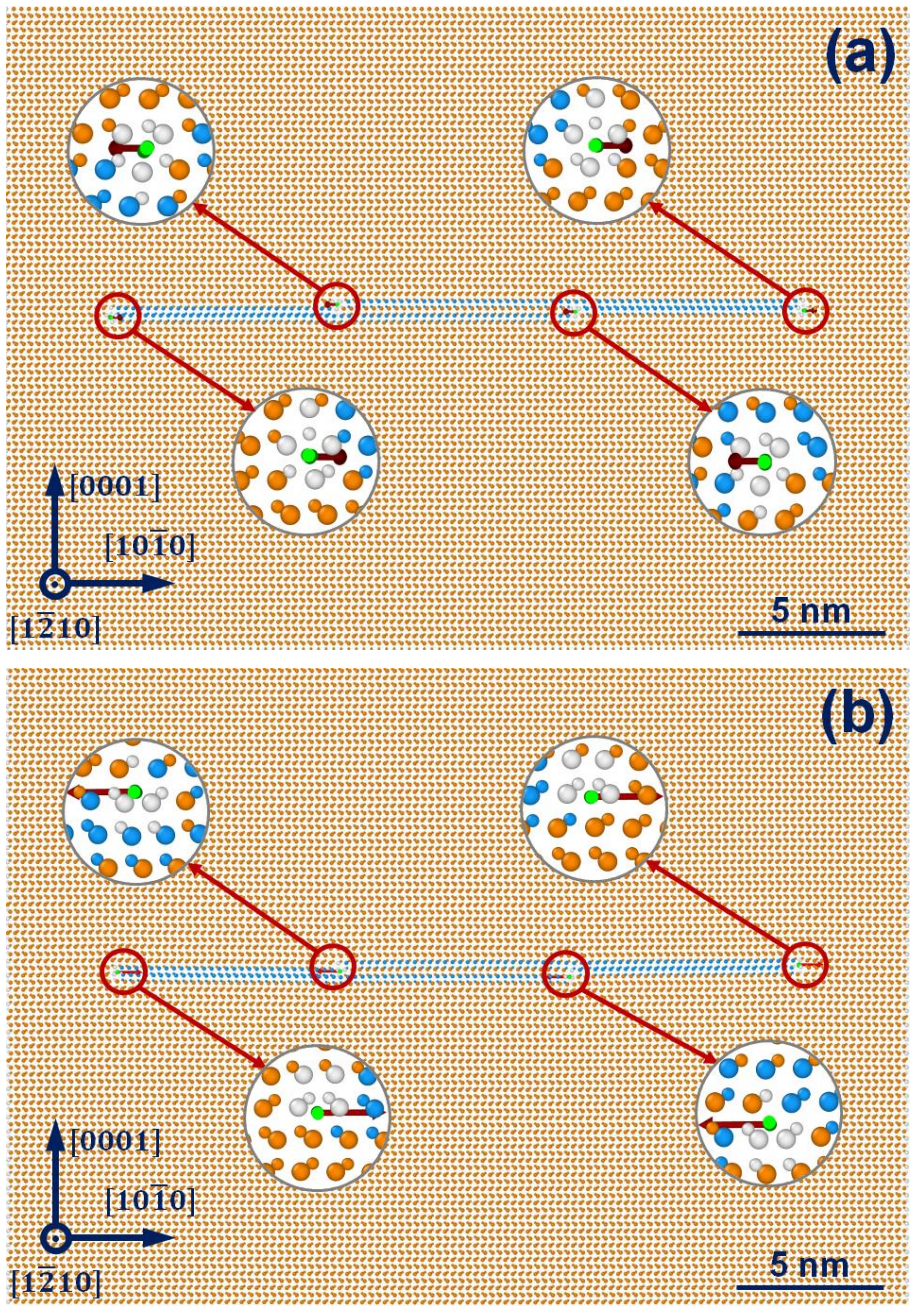

Figure 1. Initial simulation cells with 30° (a) and 90° (b) Shockley partial dislocations dipoles. The insets in the figure show enlarged atomic configurations of Shockley partial dislocations composing the dislocation dipoles.

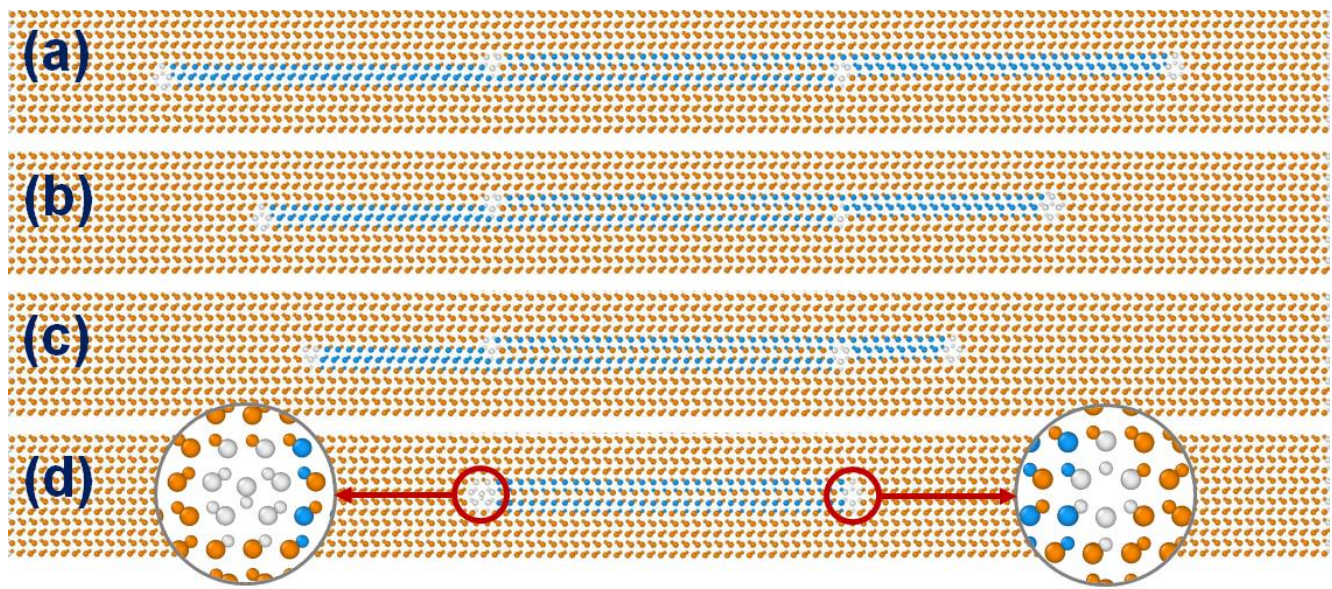

Figure 2. MD simulated evolution of the 30°-30° dislocation dipoles in wurtzite GaN. Simulated time: (a) – initial structure, (b) – 100 ps, (c) – 200 ps and (d) – 300 ps. The insets in panel (d) show enlarged atomic configurations of the formed stable dislocation complexes with zero total Burgers vectors.

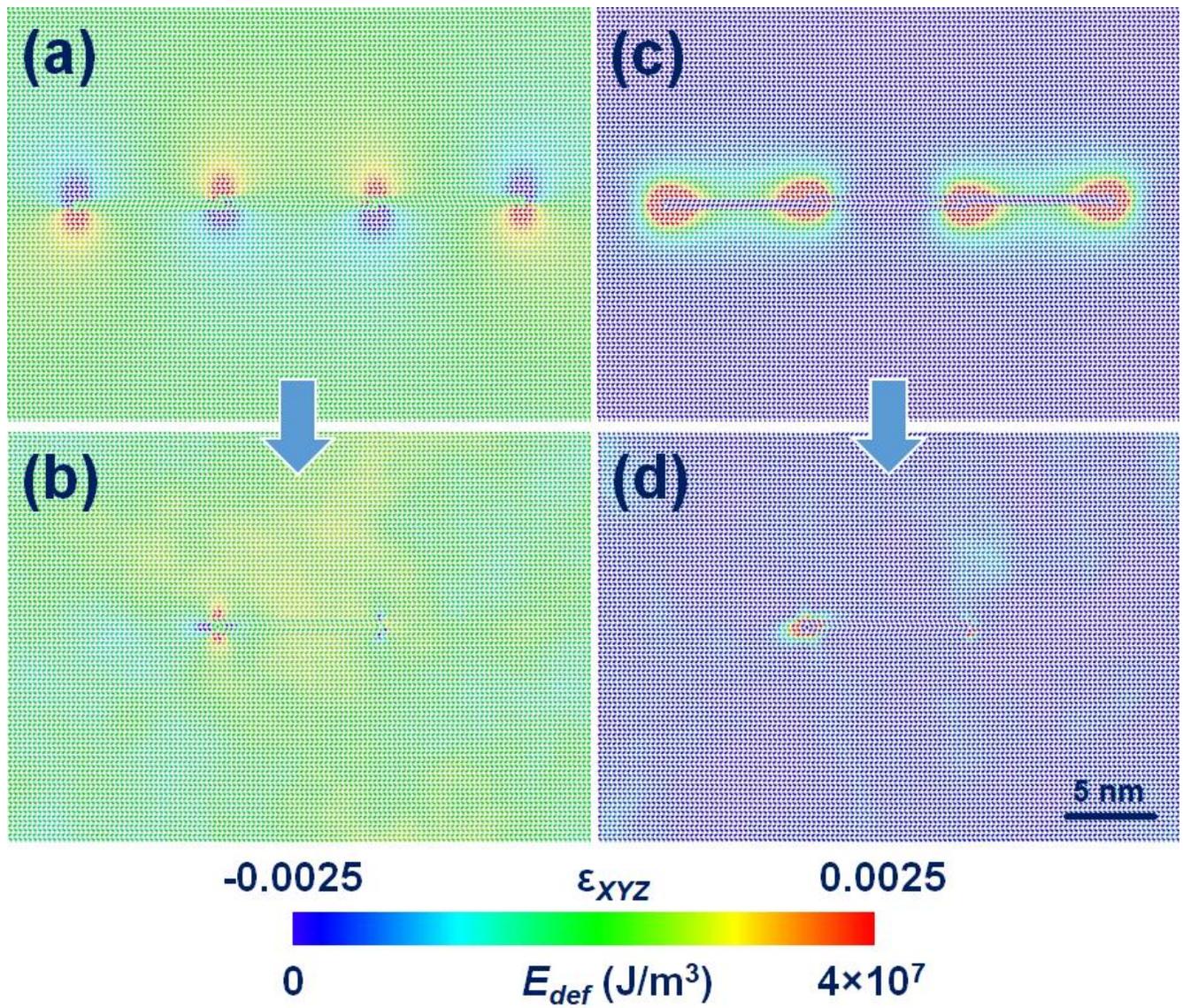

Figure 3. Distribution of volumetric strain (a, b) and elastic energy density (c, d) in the wurtzite GaN cell with 30°-30° dislocation dipoles before (a, c) and after (b, d) formation of double partial dislocation complexes.

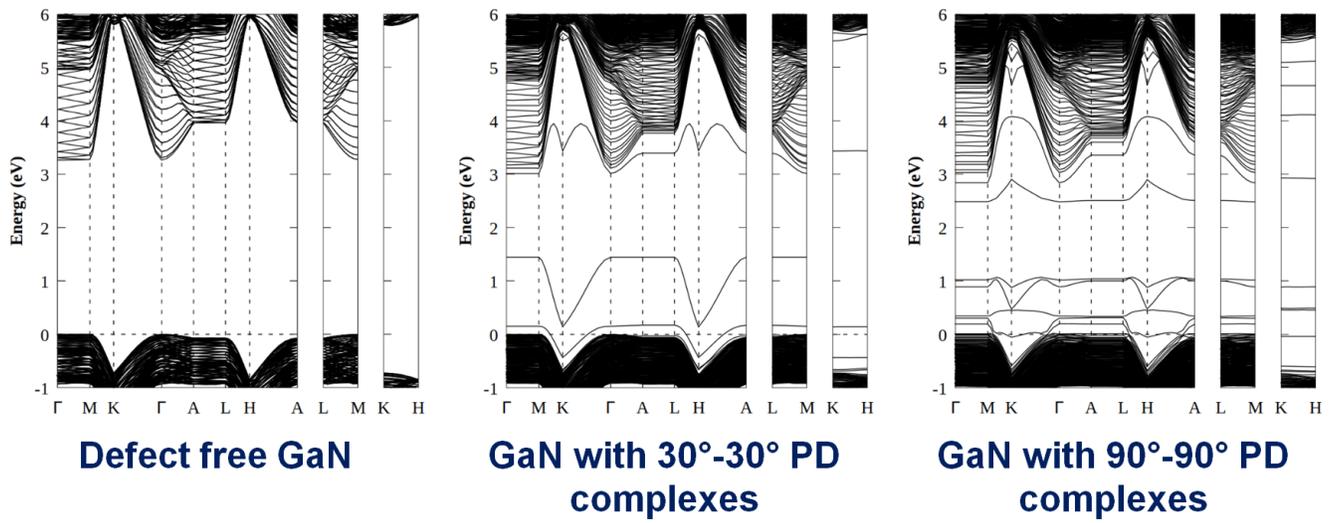

Fig. 4. DFT calculated band structure of the supercells of defect free GaN as well as GaN containing 30°-30° and 90°-90° partial dislocation (PD) complexes.

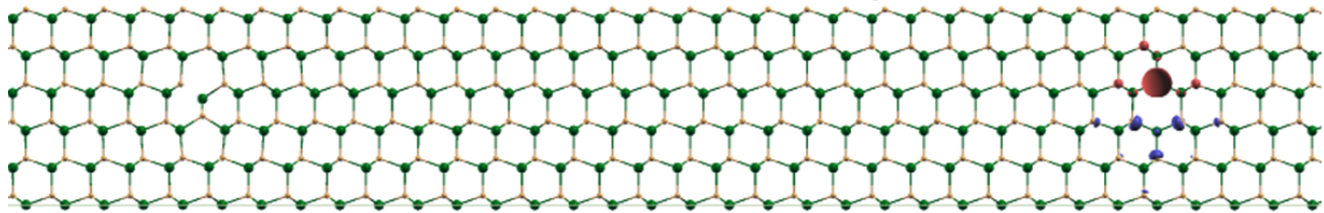

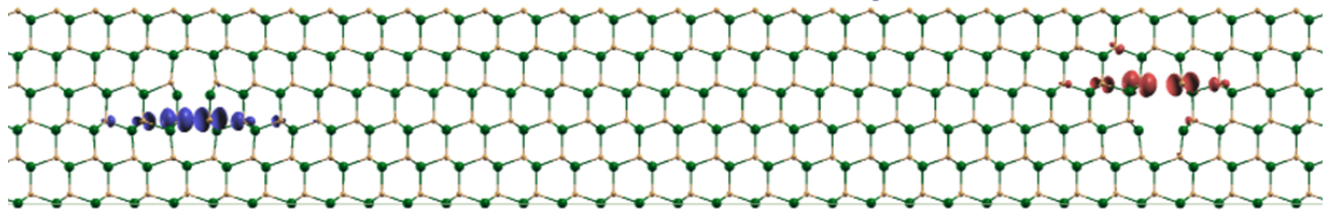

Fig. 5. Distribution of the wave functions corresponding to the minimum and maximum of the defect zones (HOMO and LUMO levels) for the supercells with 30°-30° (top panel) and 90°-90° (bottom panel) partial dislocation complexes. Blue color corresponds to HOMO and red color to LUMO.

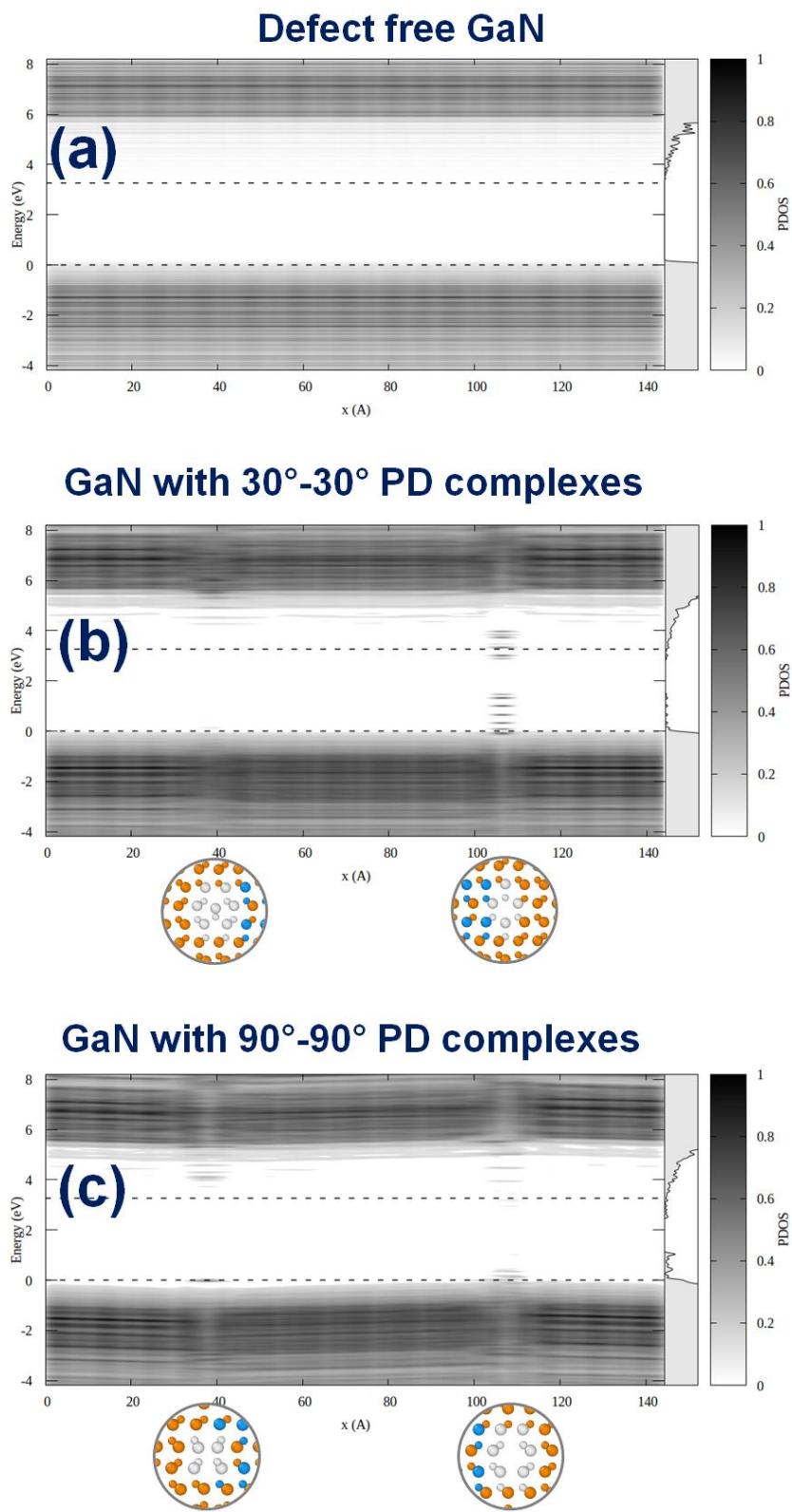

Fig. 6. Partial density of states for the supercells of defect free GaN (a) as well as GaN with 30°-30° (b) and 90°-90° (c) partial dislocation (PD) complexes.

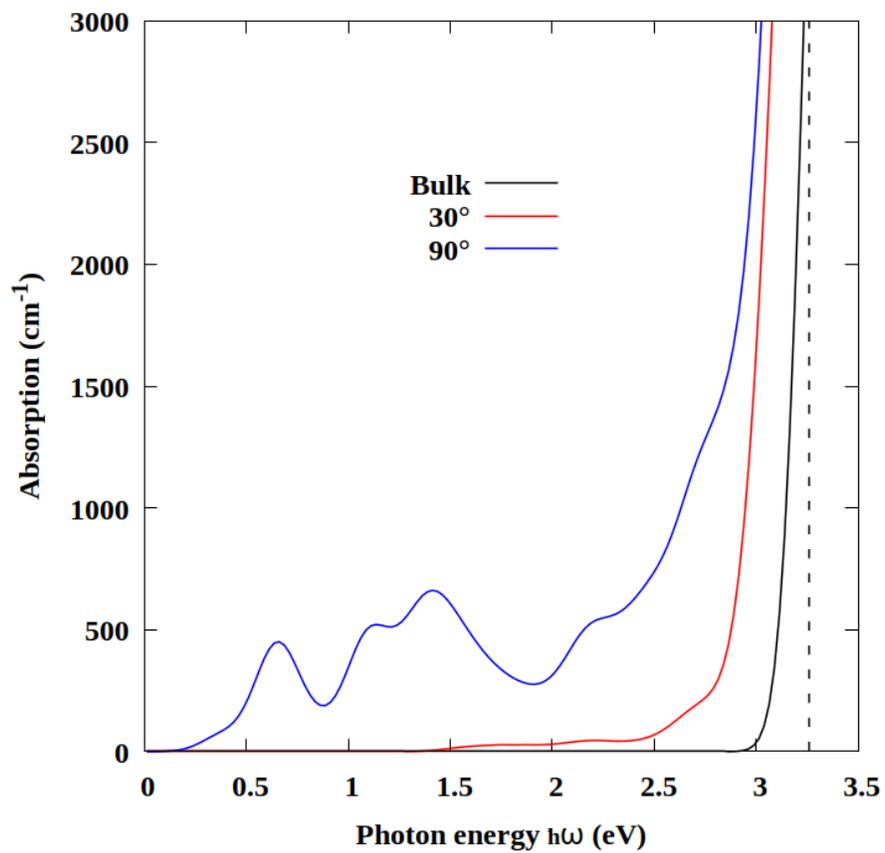

Fig. 7. DFT calculated absorption spectra for the supercells of defect free GaN as well as GaN with 30°-30° and 90°-90° partial dislocation complexes.

# Supplementary information for

# Stable partial dislocation complexes in GaN as charge carrier lifetime modifiers for terahertz device applications by molecular dynamics and first-principle simulations

Andrey Sarikov[1,2,3,*] and Ihor Kupchak[1,4]

[1] V. Lashkaryov Institute of Semiconductor Physics, National Academy of Sciences of Ukraine, 41 Nauky Avenue, 03028 Kyiv, Ukraine

[2] Educational Scientific Institute of High Technologies, Taras Shevchenko National University of Kyiv, 4-g Hlushkova Avenue, 03022 Kyiv, Ukraine

[3] National Technical University of Ukraine "Igor Sikorsky Kyiv Polytechnic Institute", 37 Beresteiskyi Avenue, 03056 Kyiv, Ukraine

[4] University of Rome "Tor Vergata", Via della Ricerca Scientifica 1, 00133 Rome, Italy

* Corresponding author. E-mail: sarikov@isp.kiev.ua

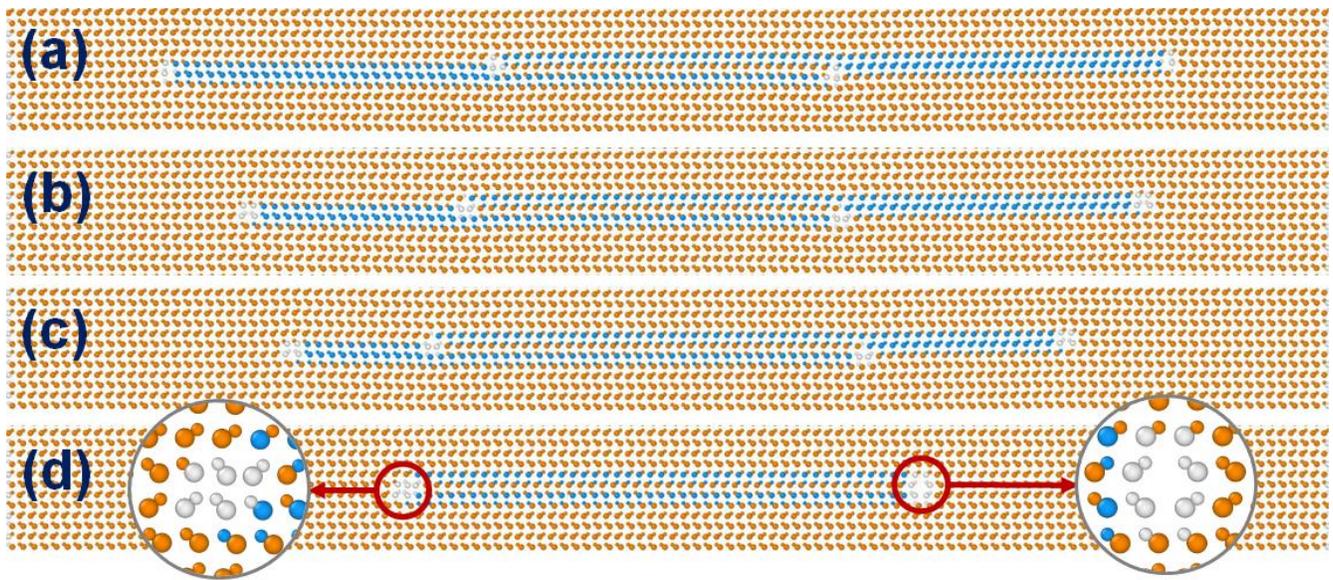

Figure 1. S. I. MD simulated evolution of the 90°-90° dislocation dipoles in wurtzite GaN. Simulated time: (a) – initial structure, (b) – 1.4 ns, (c) – 2.8 ns and (d) – 4.2 ns. The insets in panel (d) show enlarged atomic configurations of the formed stable dislocation complexes with zero total Burgers vectors.

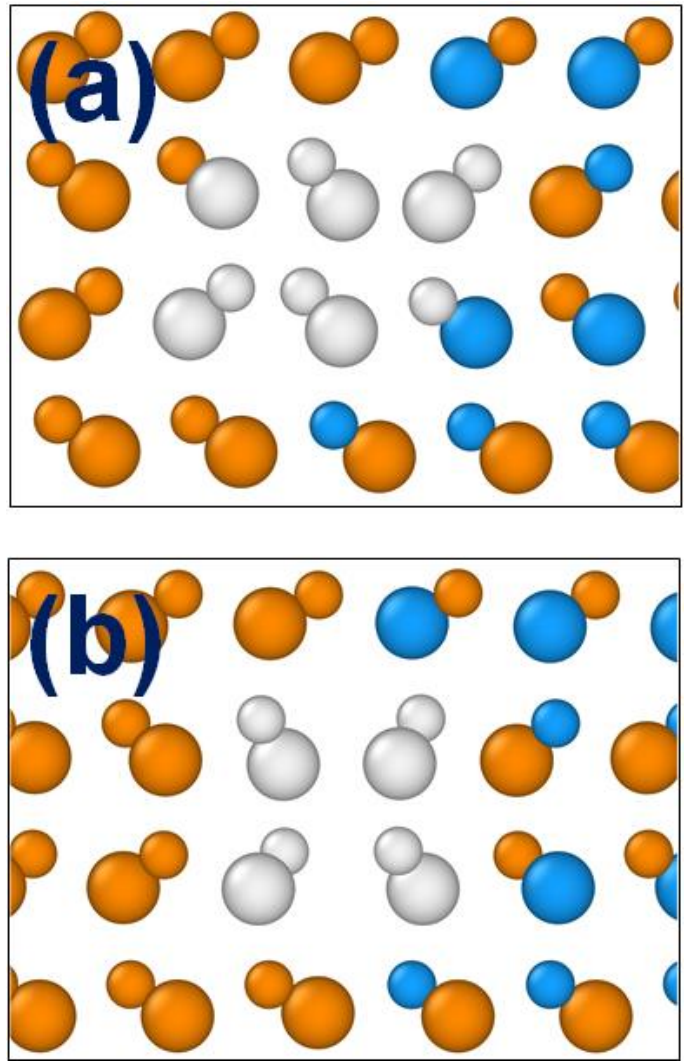

Figure 2 S. I. Atomic configuration of a flower-like 90°-90° Shockley partial dislocation complex obtained by molecular dynamics (a) and DFT (b) calculations.

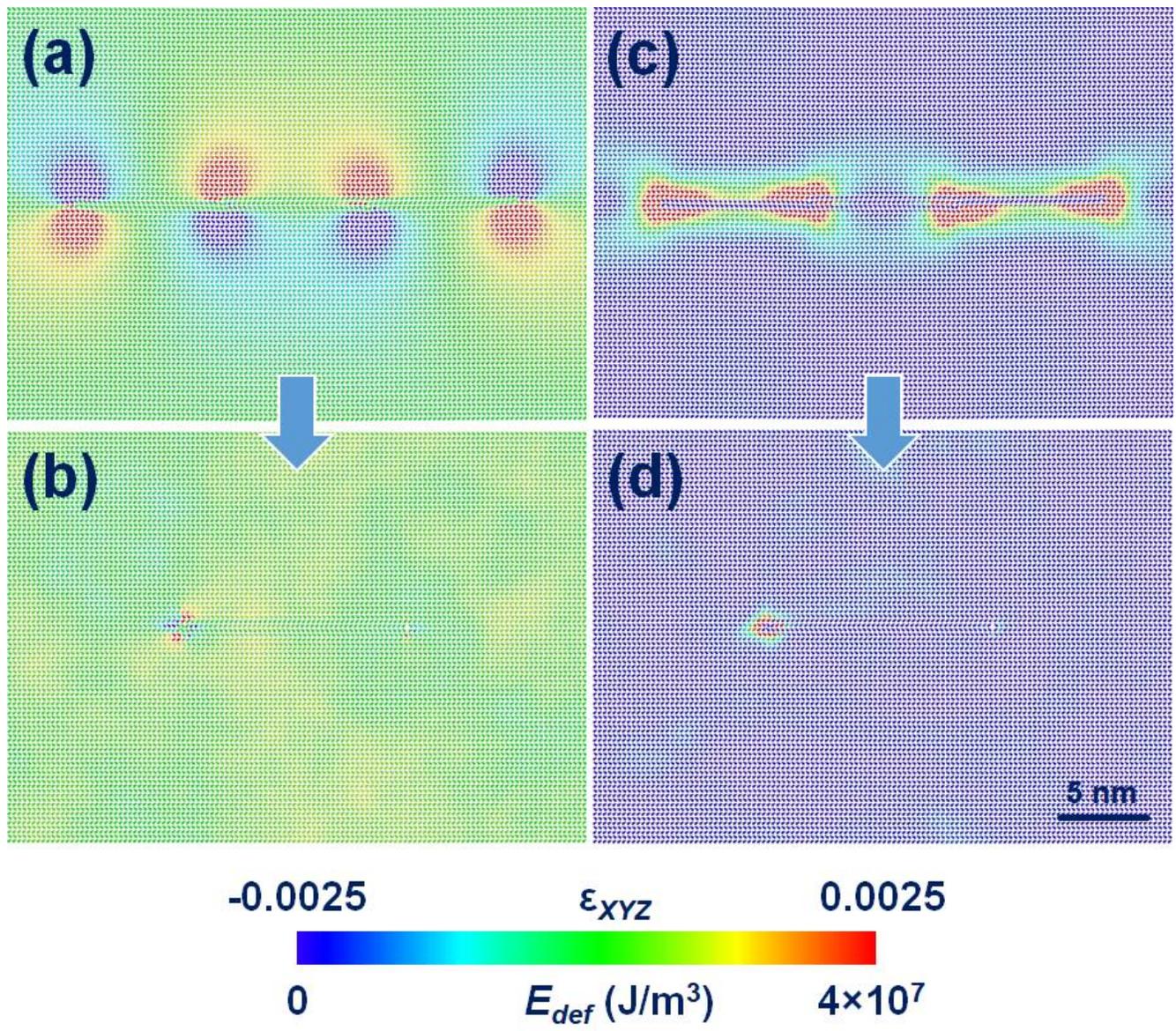

Figure 3 S. I. Distribution of volumetric strain (a, b) and elastic energy density (c, d) in the wurtzite GaN cell with 90°-90° dislocation dipoles before (a, c) and after (b, d) formation of double partial dislocation complexes.